\documentstyle[preprint,aps,eqsecnum,floats,epsf]{revtex}
\begin{document}


\draft
\preprint{\vbox{
                 \hfill AZPH-TH/95-21 \\
               }
         }

\title{Resolved Photon and Rapidity Gap in Jet Events
       \thanks{Work supported by the Department of
               Energy, contract DE--FG03--93ER--40792.
              }
       }

\author{Hung Jung Lu}

\address{Department of Physics, University of Arizona \\
         Tucson, AZ 85721 }

\date{\today}

\maketitle

\begin{abstract}
We study the production of jet events containing large rapidity
gaps in particle colliders due to the resolved photon mechanism.
In particular, we analize these events at DESY HERA and future
$e^+e^-$ linear colliders, including the possibility of $\gamma\gamma$
laser backscattering beams.
These events allow the study of the perturbative pomeron in
environments complementary to the hadron-hadron colliders,
and also provide insight into the survival probability of the
rapidity gaps involving photon initial state.
\end{abstract}


\section{Introduction}

The energy range of present and future particle colliders opens a
kinematic regime, where the observation of jet events
containing large rapidity gaps can provide interesting
physical insight into the underlying exchange mechanisms.
The presence of rapidity gaps can also serve as triggering
signal in high-mass scale physics
\cite{BJOne,BJSurvivalProbability,DokshitzerKhozeEtAl}.
Events containing large rapidity gaps have recently been
observed at hadron-hadron collider \cite{FermilabGaps}
and lepton-hadron collider \cite{HeraDISGaps,HeraPhotoproductionGaps}.
However, the possibility of observing
jets separated by large gaps is not limited to
hadron-hadron or lepton-hadron colliders. In fact,
lepton-lepton  collisions can also generate these
events \cite{EEGaps,LEPIIGaps}.
Purely leptonic collisions provide
an environment free of spectator interactions, and can give
interesting complementary information to the rapidity
gap physics of hadronic collisions.

Photon initiated collisions should provide
yet another environment for the observation of
jet events with rapidity gaps.
The initial photons could be either real
(for instance, by using a laser-backscattering
beams \cite{GinzburgKotkinSerboTelnov}), or
quasi-real, as those produced in the photoproduction events
in $ep$ and $e^+e^-$colliders via the Weis\"acker-Williams
equivalent photon mechanism \cite{WeisackerWilliams}.
In fact, in the DESY HERA case, the gap events occur both
at the deep-inelastic regime \cite{HeraDISGaps} as well as the
photoproduction regime \cite{HeraPhotoproductionGaps}.
(See also the theoretical models in Ref. \cite{HeraGapModels}.)
However, at the moment the experimental situation has focused
on the low $|t|$ region, where the proton or its low-mass
excited state (like a $\Delta$) propagates in the forward
direction, escaping detection. In this paper we explore into
the higher $|t|$ region, where the proton is broken and
a hard jet with $p_T^2 \sim |t| \gg 1$ GeV$^2$ is generated
from the broken proton. This corresponds to the double
diffractive dissociation of the $\gamma p$ system.

{}From the study of resolved photon processes
\cite{Witten,LlewellynSmith,BrodskyEtAl}
(for a review, see Ref. \cite{DreesGodbolePramana}),
we know that for the generation of jet events
with fragments in forward and backward beam direction,
the initial lepton can themselves be treated
as containing hadronic components.
Thus, one can talk about quark and gluon contents of
electrons. For these jet events, the initial leptons
would behave like supplier of the partons for the
hard subprocess, much like the initial hadrons in
hadron-hadron colliders.

It is well known \cite{DreesGodboleLinearColliders} that
as the collision energy $\sqrt{s}$ increases,
the cross section for the annihilation events at lepton
colliders decreases like $1/s$ or at best $\ln s/s$.
At the same time, the cross section for the
simplest hard two-photon process, $e^+e^- \to
e^+e^- q \bar q$ increases like $\ln^3 s$,
for fixed transverse momentum of the quarks or
fixed invariant mass of the $q\bar q$ pair.
In fact, at high enough energies $\sqrt{s} \sim
1 \ {\rm TeV}$, the combination of the increased
cross section for resolved photon processes and
the enhanced photon flux due to beamstrahlung
can lead to severe hadronic backgrounds at
$e^+e^-$ supercolliders.

The photon content of electron
is suppressed by $\alpha_{\rm em}$, but enhanced
by a logarithmic factor  $\sim \ln(s/4m_{\rm e}^2)$.
In the kinematic regime of our interest
($ \sqrt{s} > 300$ \ GeV), this means
that an electron beam can be qualitatively
visualized as carrying a photon
beam with a flux  $5 \sim 10$\% that of the electron.
Now, about a fraction $\alpha_{\rm em} \sim 10^{-2}$ of the time,
we find a parton (quark or gluon) inside a photon.
That is, taking all effects, electrons carry hadronic
partons at the $10^{-3}$ level. If we consider
double resolved processes, this means a factor of
$10^{-3} \times 10^{-3} = 10^{-6}$ reduction in the flux.
However, the parton diffractive scattering cross section is
rather large, and it is essentially controlled by the cut-off in
transverse momentum. This should be contrasted
to the $e^+e^-$ annihilation cross section, which
is controlled by the total center-of-mass
energy $\sqrt{s}$. While the annihilation cross section becomes
very small at large $\sqrt{s}$, the diffractive cross section
at fix transverse momentum cut increases with $\sqrt{s}$.
Therefore, we would expect the resolved photon diffractive
events to be produced at an reasonable level at future $e^+e^-$
colliders, despite the flux reduction.

For the real $\gamma\gamma$ colliders, we expect
cross sections much larger than the $e^+e^-$ case, because the
photons now are not coming from the electrons, hence
there is less suppression of parton flux.
( At high energies, such as is planned for the
Next Linear Collider \cite{Waikoloa},
potentially one should also consider
$W$ and $Z$ bosons as partons inside electrons. But we will
limit our scope here to the resolved photon contribution. )

For the resolved photon mechanism of generating jet events
containing large rapidity gaps, we expect the spectator
partons inside the photon to interact with the spectator
partons inside the opposite beam particle. That is, the
situation is similar to the case of hadron-hadron collider.
The soft spectator interaction can generate particles and spoil
the rapidity gap. The rapidity survival probability
$\langle S^2 \rangle$ is defined as the probability that
no other interaction occurs beside the hard interaction of interest
\cite{BJSurvivalProbability}.
This probability is most readily estimated as an average over
the hadron-hadron impact parameter $B$ \cite{BJSurvivalProbability}:
\begin{equation}
\langle S^2 \rangle
=
\frac{\int d^2 B f(B) S^2(B)}
     {\int d^2 B f(B)},
\end{equation}
where $S^2(B)$ is the probability that the colliding
hadrons do not interact inelastically, and $f(B)$ is
the cross section for the hard collision of interest.
Different estimates for $\langle S^2 \rangle$ in
hadron collisions, based on a variety of phenomenological
models, are presented in Ref.
\cite{GotsmanLevinMaor}, where $\langle S^2 \rangle$ is
estimated to be between $0.05$ and $0.2$.
$\langle S^2 \rangle$ is expected to depend on the colliding
energy, but only weakly on the size of the rapidity gap.
In Ref. \cite{Fletcher} the author uses a Good-Walker model
for diffraction, and obtains a much higher value for the
survival probability ($44\%$ at Tevatron energies, and
$33\%$ at $40$ TeV.) On a related issue, the authors in
Ref. \cite{ButterworthEtAl} have used the HERWIG Monte Carlo
program and found that in $\gamma\gamma$ and $\gamma p$
collisions, the mean number of hard interactions per event
ranges from $1.04$ to $1.123$ for various particle colliders.

To obtain the cross section of jet events with rapidity
gaps, we must multiply the hard collision cross section
by the survival probability, that is,
\begin{equation}
\sigma_{\rm gap} = \langle S^2 \rangle \sigma_{\rm hard}.
\end{equation}
Qualitatively, we expect the survival probability
involving photon initial states to be approximately the
same of the survival probability involving only hadronic
initial states. This can be argued based on the vector meson dominance
picture. However, it would be very interesting to study
the difference of survival probabilities from hadronic
and photonic initial states. The measurement of resolved photon
gap events can clarify this difference.

In this paper we will present only the hard cross sections,
without taking into the soft physics of the survival
probability. But it will be implicitly understood that
in order to obtain the final cross sections, the factor
$\langle S^2 \rangle$ must be multiplied.

This paper is structured as follows.
In Section II we will study the gap event cross section at HERA
in the photon-proton double diffractive dissociation region.
In Section III we analyze the
situation at future linear collider energies ($0.5$ to $1.5$ TeV),
and in Section IV we analyze the situation for $\gamma\gamma$
colliders for a similar region of energies. Finally, in Section
V we give the conclusions.

\section{Resolved Photon Gap Events at HERA}

The mechanism of generating resolved photon
gap events at HERA $ep$ collider is illustrated in Fig. 1.
The partons that participate in the hard collision
subprocess can be either quarks or gluons. We will
consider the hard collision regime $|t| \gg 1$ GeV$^2$ where
the perturbative pomeron is applicable. Notice that this
kinematic regime differs from the previous gap event regimes
in Ref. \cite{HeraDISGaps,HeraPhotoproductionGaps}, where
the proton remains unbroken or is excited to a low-mass
state, and propagates down the beam pipe, escaping detection.
That is, the rapidity gaps observed so far in HERA are
between the real or virtual photon fragments and the
unbroken proton or its excited state. This situation is
different from the gap events observed at hadron-hadron
colliders \cite{FermilabGaps}, where the rapidity gap is
observed between two measured jets. Here in our paper,
we consider a situation similar to the hadron-hadron
collision case. That is, hard jets are generated both from
the photon fragmentation and the proton fragmentation regions.
This corresponds to the double diffractive dissociation of
the $\gamma p$ system, and the rapidity gap exists between
the two observed hard jets.

\begin{figure}[htbp]
\begin{center}
\leavevmode
{
 \epsfxsize=3.00in
 \epsfbox{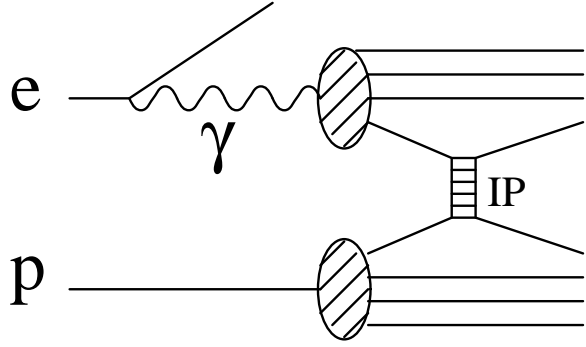}
}
\end{center}
\caption[*]{
            Resolved photon mechanism for producing jet
            events with a large rapidity gap in $ep$
            collision. The partons inside the photon and
            the proton undergo a hard scattering via the
            exchange of a perturbative QCD pomeron.
           }
\label{Fig1}
\end{figure}

The diffractive scattering cross section
for the quark-quark $t$-channel color-singlet exchange case
has been obtained by Mueller and Tang \cite{MuellerTang}
(see also Ref. \cite{MorePomeron}) by using the
Balitsky-Fadin-Kuraev-Lipatov (BFKL) model
\cite{BFKLPomeron}:
\begin{equation}
\frac{d\sigma_{qq}}{dt} =
\bigl(\alpha_{\rm s} C_{\rm F}
\bigr)^4
\frac{\pi^3}{4t^2}
\frac{e^{2(\alpha_{\rm P}-1)y}}
     { \Bigl[ \frac{7}{2} \alpha_{\rm s}
              C_{\rm A} \zeta (3) y
       \Bigr]^3
     },
\end{equation}
where $\alpha_{\rm s}=\alpha_{\rm s}(-t)$
is the strong coupling constant,
$y= \ln(-\hat{s}/t)$ is the rapidity interval
between the two out-going partons as measured
from the beam axis, where $\hat{s}$ is the
total center-of-mass energy squared of the
$qq$ system;
$\alpha_{\rm P} = 1+(4\alpha_{\rm s}C_{\rm A}/\pi) \ln 2$ the slope
of the pomeron trajectory, $\zeta(3)=1.20206\dots$, and
$C_{\rm A} = 3$ and $C_{\rm F} = 4/3$
the values of the Casimir operators in the adjoint and fundamental
representations of the $SU(3)$ group.
For the case of gluon-gluon elastic scattering
with color-singlet $t$-channel exchange, we need only to replace
the $C_{\rm F}$ factor in $d\sigma_{qq}/dt$ by $C_{\rm A}$. That is,
\begin{equation}
\frac{d\sigma_{gg}}{dt} =
\alpha_{\rm s} C_{\rm A}
\frac{\pi^3}{4t^2}
\frac{e^{2(\alpha_{\rm P}-1)y}}
     { \Bigl[ \frac{7}{2} \zeta (3) y
       \Bigr]^3
     }.
\end{equation}

To obtain the total cross section, we must integrate the
cross section of the hard collision weighed by the
respective structure functions.

\begin{eqnarray}
\sigma_{ep} (s,m_{\rm cut}^2,Y_{\rm cut})
&=&
\int_{(m_{\rm cut}^2, Y_{\rm cut})}
dz dx_1 \ dx_2 \ dt \
f_{\gamma|e} (z,s)
\cr
& &
P_{\gamma}(x_1,-t) \ P_p(x_2,-t) \
\frac{d\sigma_{gg}}{dt}
\bigl( \hat{s} = z x_1 x_2 s, t
\bigr).
\label{EqHeraGapCrossSection}
\end{eqnarray}
In the above formula, $z$ is the momentum fraction of the
incoming electron carried by the photon, $x_1$ and $x_2$
are the momentum fraction of the partons carried by
the photon and proton, respectively.
$\sqrt{s}$ is the total center-of-mass energy of the
$ep$ system, $m_{\rm cut}^2$ is the minimum transverse
momentum squared of the hard tagging jets,
and $Y_{\rm cut}$ is the minimum rapidity interval separating
the two hard jets.
The number of photons carrying a fraction $z$
of the energy of an emitting electron in leading
log approximation is given by
$f_{\gamma|{\rm e}}(s,z)=
 \alpha_{\rm em}/(2\pi z)
 [1+(1-z)^2] \ln(s/4m_{\rm e}^2)
$, with $\sqrt{s}$ the total center-of-mass energy
of the colliding $ep$ system
and $m_{\rm e}$ the electron mass
\cite{WeisackerWilliams,BawaStirling}.
However, this formula overestimates the direct
$\gamma\gamma \to q \bar{q}$ contribution
\cite{BhattacharyaSmithGrammer}.
An improved expression including non-leading terms
is \cite{BrodskyKonishitaTerazawa}:
\begin{eqnarray}
f_{\gamma|{\rm e}}(z,s)=
\frac{\alpha_{\rm em}}{2 \pi z}
&\Biggl\{& \Bigl[ 1 + (1-z)^2
           \Bigr]
           \Bigl( \ln(s/4m_{\rm e}^2) -1
           \Bigr)
\cr
&+&   z^2 \Biggl[ \ln \frac{2 (1-z)}{z} + 1
          \Biggr]
 + (2-z)^2 \ln \frac{2 (1-z)}{2-z}
\Biggr\},
\end{eqnarray}
This formula has been shown
\cite{BhattacharyaSmithGrammer} to give accurate
results not only for total (direct) jet rates but
also for distributions.
In Eq. (\ref{EqHeraGapCrossSection}), the quantities
$P_{\gamma}(x_1,-t)$ and $P_p(x_2,-t)$ correspond
to the parton structure functions of the photon and
proton, respectively, and they are defined as
\begin{eqnarray}
P_{\gamma}(x,Q^2)
&=&
f_{g|\gamma} (x,Q^2)
+
\Biggl( \frac{C_{\rm F}}{C_{\rm A}}
\Biggr)^2
\sum_q
\Bigl[ \ f_{q|\gamma} (x,Q^2)
     + f_{\bar{q}|\gamma} (x,Q^2) \
\Bigr],
\cr
P_p(x,Q^2)
&=&
f_{g|p} (x,Q^2)
+
\Biggl( \frac{C_{\rm F}}{C_{\rm A}}
\Biggr)^2
\sum_q
\Bigl[ \ f_{q|p} (x,Q^2)
     + f_{\bar{q}|p} (x,Q^2) \
\Bigr].
\end{eqnarray}
For the photon structure functions, we have the identity
$f_{q|\gamma} (x,Q^2)= f_{\bar{q}|\gamma} (x,Q^2)$.
There exist many parametrizations for the
parton distributions inside proton and photon.
For the photon distribution functions, we will limit
ourselves to the Drees-Godbole (DG) parametrization
\cite{DreesGodboleParametrization}
and to the Gl\"uck-Reya-Vogt (GRV-LO) parametrization
\cite{GluckReyaVogtParametrization}. (We thank the
authors for providing the programs.)
Numerically, the largest uncertainty in our calculation
comes from the gluon density inside photon, where
little experimental result is available.
The quark densities are better understood. See the
recent results by the TRISTAN's collaborations
TOPAZ \cite{TOPAZ} and AMY \cite{AMY} on the measurement
of the photon structure function $F_2^\gamma(x,Q^2)$.
It is worth mentioning that the gluon density of the
photon in the GRV-LO parametrization
is consistent with the
recent measurement by the H1 Collaboration
at HERA \cite{H1PhotonStructure}.
For the proton structure function, we choose the
CTEQ 2'M parametrization from the CERN PDFLIB
routine library \cite{PlothowBesch}.
We use $Q^2=-t$
for the scale of the photon and proton
structure functions.

We perform the numerical integration with the Monte Carlo
integration program VEGAS \cite{Vegas}.
The integration limits are
\begin{eqnarray}
&& 0 \le z, x_1, x_2 \le 1, \cr
&& m_{\rm cut}^2 \le |t| \le x_1 x_2 e^{-Y_{\rm cut}} s,
\end{eqnarray}
where we integrate over all events with jet transverse momentum
larger than $m_{\rm cut}^2$ and a rapidity separation between the jet
centers larger than $Y_{\rm cut}$. It should be pointed out that due to
the hadronization effect, the hadron fragments typically
scatter around the jet centers, within a circle of radius
$\sim 0.7$ units of rapidity \cite{BJSurvivalProbability}.
Hence the observed effective gap is $Y_{\rm eff} \sim Y - 1.4$.

\begin{figure}[htbp]
\begin{center}
\leavevmode
{
 \epsfysize=5.00in
 \epsfbox{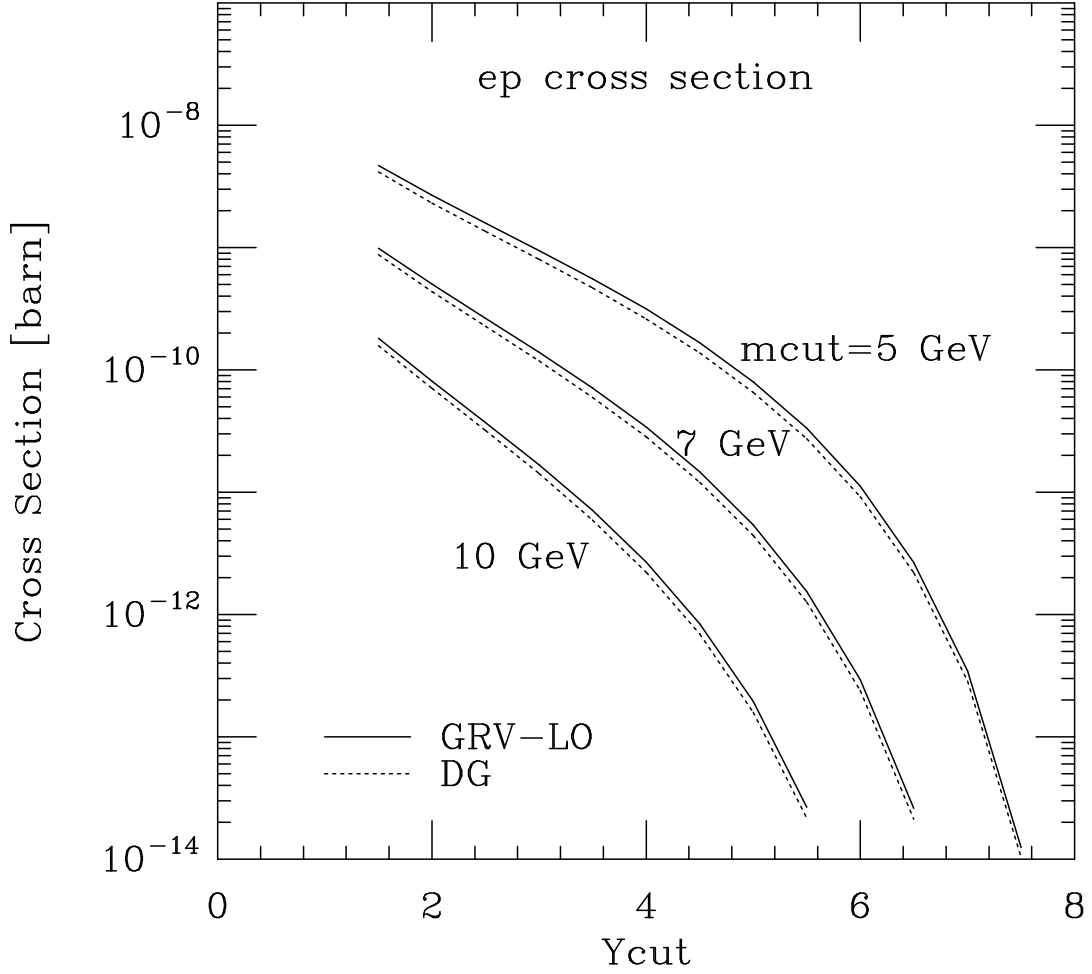}
}
\end{center}
\caption[*]{
            Resolved photon gap event cross section at HERA,
            for various values of the transverse moment cut
            $m_{\rm cut}$. The solid lines are obtained with
            the Gl\"uck-Reya-Vogt parametrization of the
            photon structure functions, and the  broken lines
            with the Drees-Godbole parametrization.
           }
\label{Fig2}
\end{figure}

In Fig. 2 we present the result of the calculation.
For $Y_{\rm cut}=4$ and $m_{\rm cut}=5$ \ GeV, even taken into account
the survival probability consideration and uncertainty
for the gluon distribution inside the photon, the
gap events should still be produced at an observable
rate. As we stated before, gap events have been observed
at HERA both in the deep-inelastic regime \cite{HeraDISGaps}
and in the photoproduction regime \cite{HeraPhotoproductionGaps},
where the rapidity gap exists between the photon (real
or virtual) fragmentation region and the forward,
undetected proton (or its excited state.)
It would be interesting to verify the existence of
rapidity gap between two observed hard jets at HERA, and
study the dependence of these gap event
cross sections on the rapidity interval $Y_{\rm cut}$ and on the
transverse moment cut $m_{\rm cut}^2$, and compare with our
calculation here. We should note, however, that smaller rapidity
gaps can also arise from random fluctuation of the fragments of
hadronization process. A systematic study of rapidity gap
physics here and in other environments \cite{FermilabGaps} will
gradually allow better understanding of the relative
importance of gap events from perturbative mechanisms and
from random fluctuations, as well as insight to
the survival probability involving photons.

It is important to point out, though, that our current
knowledge of the photon structure functions is rather
imprecise. This is especially true for the gluon content
of photon, which numerically forms the dominant contribution
to the gap event cross section. Therefore, an uncertainty
of half an order of magnitude above or below the calculated curves
would not be unreasonable. Hopefully the gluon content of
the photon can be better measured at HERA in the future
\cite{H1PhotonStructure}.

\section{Resolved Photon Gap Events at NLC}

\begin{figure}[htbp]
\begin{center}
\leavevmode
{
 \epsfxsize=3.00in
 \epsfbox{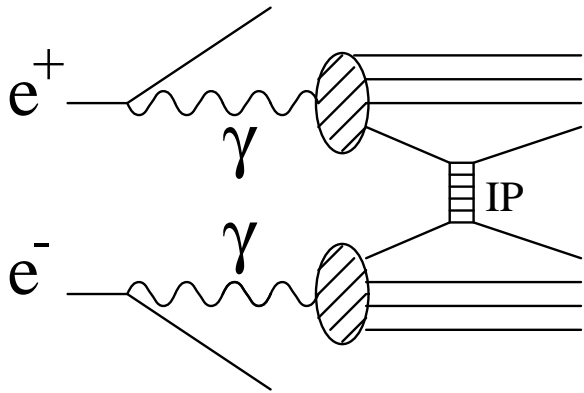}
}
\end{center}
\caption[*]{
            Resolved photon mechanism for producing jet
            events with a large rapidity gap in $e^+e^-$
            collision. The partons inside the photons
            undergo a hard scattering via the exchange
            of a perturbative QCD pomeron.
           }
\label{Fig3}
\end{figure}

The production mechanism is illustrated in Fig. 3.
Currently, the energy projected for future $e^+e^-$
colliders is in the range $0.5$ to $1.5$ TeV. At
this energy range, the resolved photon contributions
should be substantial. However, due to beamstrahlung
background, the detectors for these colliders are
not expected to be sensitive to jet production in the forward
or backward direction.
That is, many of the gap events will not be detected.
We will analyze the production cross section
of gap event by taking
into account also the detector limitation. Here we will
assume that the detector is only capable of observing hard jets
produced in the rapidity region
$[-\eta_{\rm det}, \eta_{\rm det}]$.
Given this limitation, the observable
rapidity gap events can be classified into three cases:
(a) both hard jets are observed, (b) only the forward hard jet is
observed, and (c) only the backward hard jet is observed.
These situations are illustrated in Fig. 4. In the cases (b) and
(c), since one of the hard jets is not measured, it is not
possible to know the true size of the rapidity gap. For these
cases, we define the empirical rapidity gap as the size of the
rapidity interval between the hard jet and the detector limit
on the opposite side. Mathematically, if $y_1$ and $y_2$ represent
the rapidities of the forward and the backward hard jet, then
we define the empirical gap to be
\begin{equation}
Y = {\rm Min}
\left\{ y_1 - y_2,
        y_1 + \eta_{\rm det},
        \eta_{\rm det} - y_2
\right\} ,
\end{equation}
Naturally, the size of the rapidity gap cannot exceed
the detector range: $Y < 2 \eta_{\rm det}$.

\begin{figure}[htbp]
\begin{center}
\leavevmode
{
 \epsfxsize=4.00in
 \epsfbox{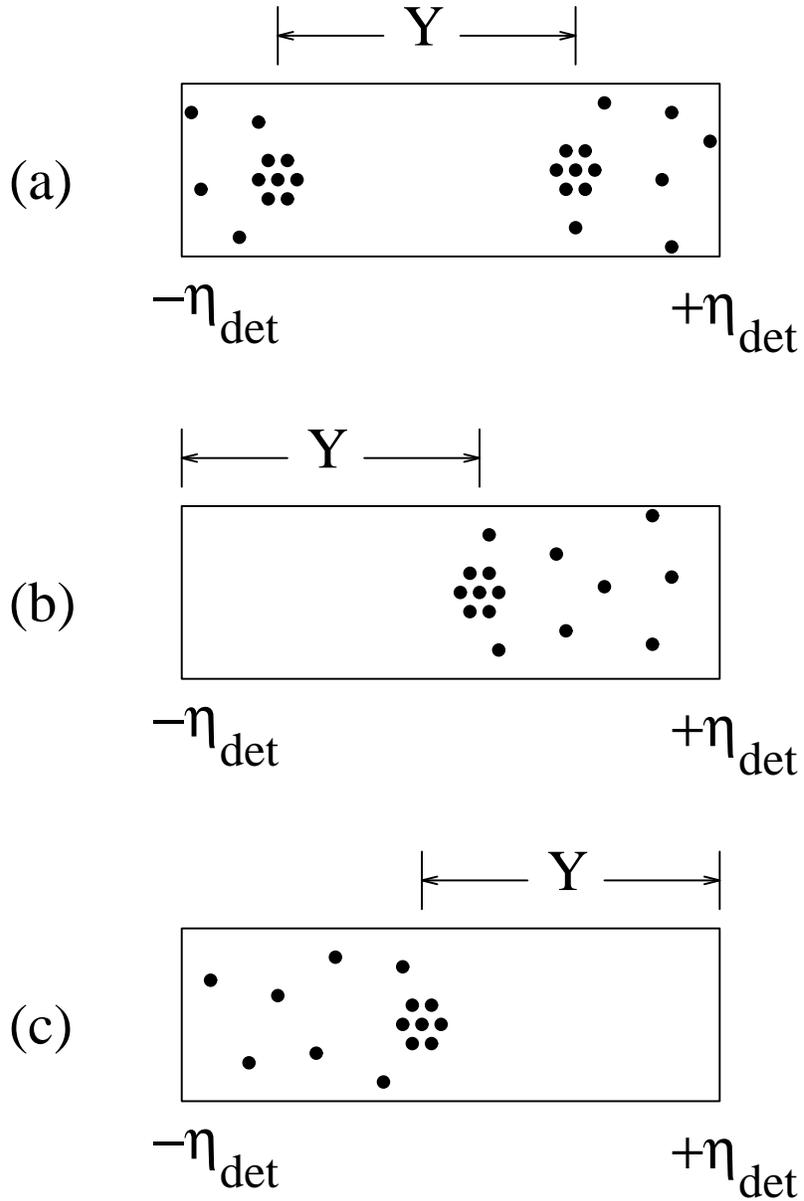}
}
\end{center}
\caption[*]{
            Three possible cases for the observation
            of jet events containing rapidity gaps:
            (a) both hard jets are observed, (b) only
            the forward jet is observed, and (c) only
            the backward jet is observed. In the cases
            of (b) and (c), the gap size is measured from
            the jet center to the rapidity edge of the
            detector on the opposite side.
           }
\label{Fig4}
\end{figure}

To integrate the event cross section, we take into account
the kinematic cuts from 1) detector limitation: $\eta_{\rm det}$,
2) rapidity gap cut: $Y_{\rm cut}$,
and 3) transverse momentum cut: $m_{\rm cut}^2$.
\begin{eqnarray}
\sigma_{e^+e^-} (s,m_{\rm cut}^2,Y_{\rm cut},\eta_{\rm det})
&=&
\int_{(m_{\rm cut}^2, Y_{\rm cut}, \eta_{\rm det})}
dz_1 dz_2 dx_1 \ dx_2 \ dt \
f_{\gamma|e} (z_1,s)
f_{\gamma|e} (z_2,s)
\cr
& &
P_{\gamma}(x_1,-t) \ P_{\gamma}(x_2,-t) \
\frac{d\sigma_{gg}}{dt}
\bigl( \hat{s} = z_1 z_2 x_1 x_2 s, t
\bigr).
\label{EqNLCGapCrossSection}
\end{eqnarray}
The photon momentum fractions inside the electron and
the positron are respectively $z_1$ and $z_2$. Other
quantities that appear in this formula have
been explained in the previous section. The integration
momentum fractions are constrained to:
$0 \le z_1, z_2, x_1, x_2 \le 1.$
The gap and transverse momentum constraints impose
the following limits for the $t$ integration. For the
case (a),
\begin{equation}
{\rm Max}
\left\{
       \begin{array}{c}
        m_{\rm cut}^2 \\
        x_1^2 z_1 z_2 e^{-2\eta_{\rm det}} s \\
        x_2^2 z_1 z_2 e^{-2\eta_{\rm det}} s \\
       \end{array}
\right\}
\ < \ \left| t \right|
\ < \
x_1 x_2 z_1 z_2 e^{-Y_{\rm cut}} s .
\label{EqLimitA}
\end{equation}
For the case (b),
\begin{equation}
{\rm Max}
\left\{
       \begin{array}{c}
        m_{\rm cut}^2 \\
        x_1^2 z_1 z_2 e^{-2\eta_{\rm det}} s \\
       \end{array}
\right\}
\ < \ \left| t \right|
\ < \
{\rm Min}
\left\{
       \begin{array}{c}
        x_1^2 z_1 z_2 e^{-2(Y_{\rm cut}-\eta_{\rm det})} s \\
        x_2^2 z_1 z_2 e^{-\eta_{\rm det}} s
       \end{array}
\right\} .
\label{EqLimitB}
\end{equation}
And for the case (c),
\begin{equation}
{\rm Max}
\left\{
       \begin{array}{c}
        m_{\rm cut}^2 \\
        x_2^2 z_1 z_2 e^{-2\eta_{\rm det}} s \\
       \end{array}
\right\}
\ < \ \left| t \right|
\ < \
{\rm Min}
\left\{
       \begin{array}{c}
        x_1^2 z_1 z_2 e^{-\eta_{\rm det}} s \\
        x_2^2 z_1 z_2 e^{-2(Y_{\rm cut}-\eta_{\rm det})} s \\
       \end{array}
\right\} .
\label{EqLimitC}
\end{equation}
We include all three contributions in our integration of
the event cross sections. In Fig. 5 and Fig. 6 are the
results of the gap event cross sections, for $0.5$ and
$1.5$ TeV center-of-mass energy. We consider three values
for the maximum detector rapidity:
$\eta_{\rm det} = 2, 3, 4$, which means that the maximum
detectable rapidity gaps are respectively $Y=4, 6, 8$.
The curves are plotted for two values of the transverse
momentum cut: $m_{\rm cut} = 5, 10$ GeV.

\begin{figure}[htbp]
\begin{center}
\leavevmode
{
 \epsfysize=5.00in
 \epsfbox{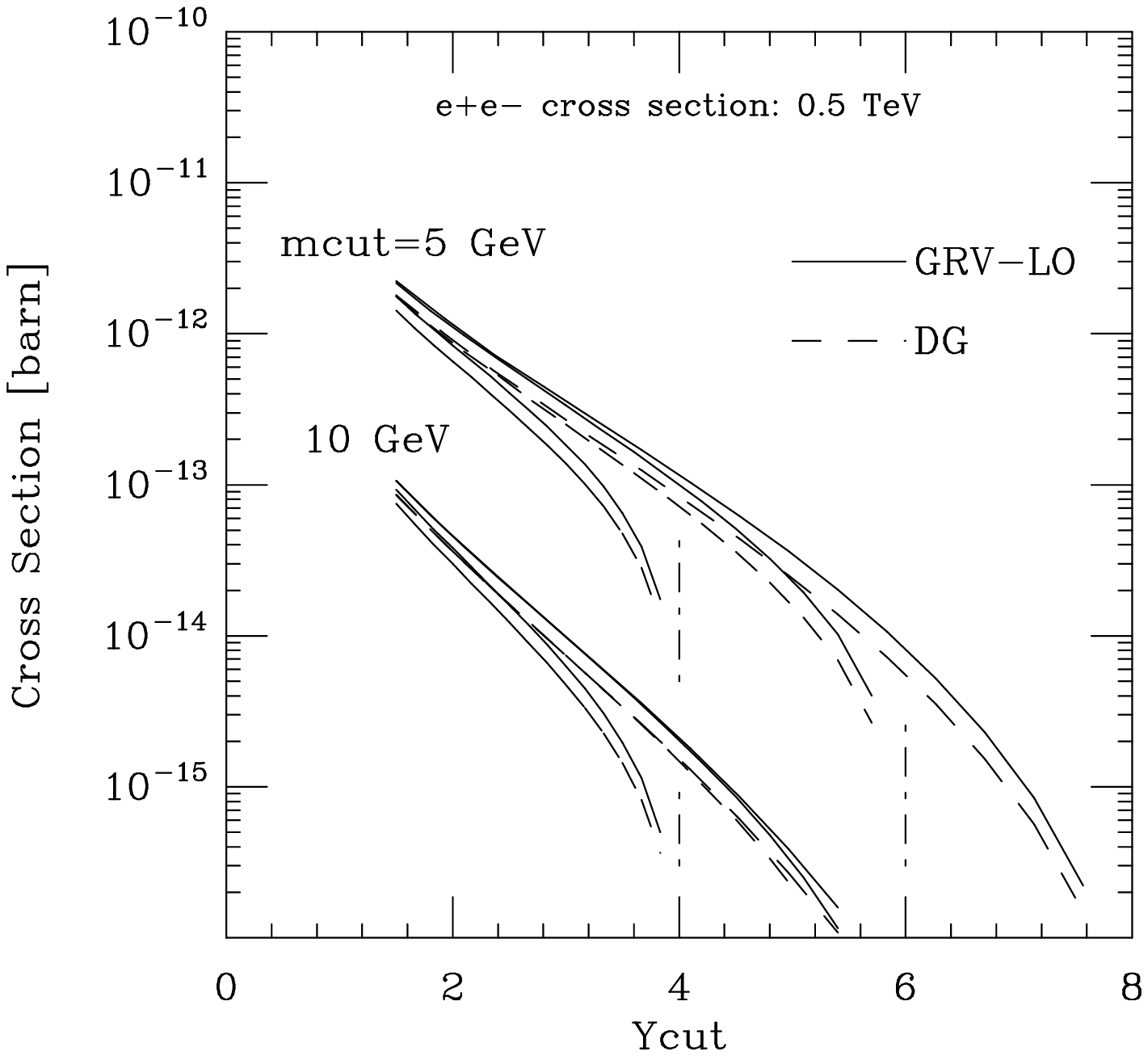}
}
\end{center}
\caption[*]{
            Gap event cross section for $e^+e^-$ collider
            at $E_{\rm cm}=0.5$ TeV, for various values
            of transverse momentum cut $m_{\rm cut}=5,10$ GeV
            and detector rapidity limits $\eta_{\rm det}=2,3,4$.
            The maximum observable rapidity gaps are
            $2 \eta_{\rm det} = 4,6,8$ and are indicated by the
            dot-dashed vertical lines. The solid lines are
            obtained by using the Gl\"uck-Reya-Vogt parametrization
            of the photon structure functions, and the broken lines
            by using the Dree-Godbole parametrization.
           }
\label{Fig5}
\end{figure}

\begin{figure}[htbp]
\begin{center}
\leavevmode
{
 \epsfysize=5.00in
 \epsfbox{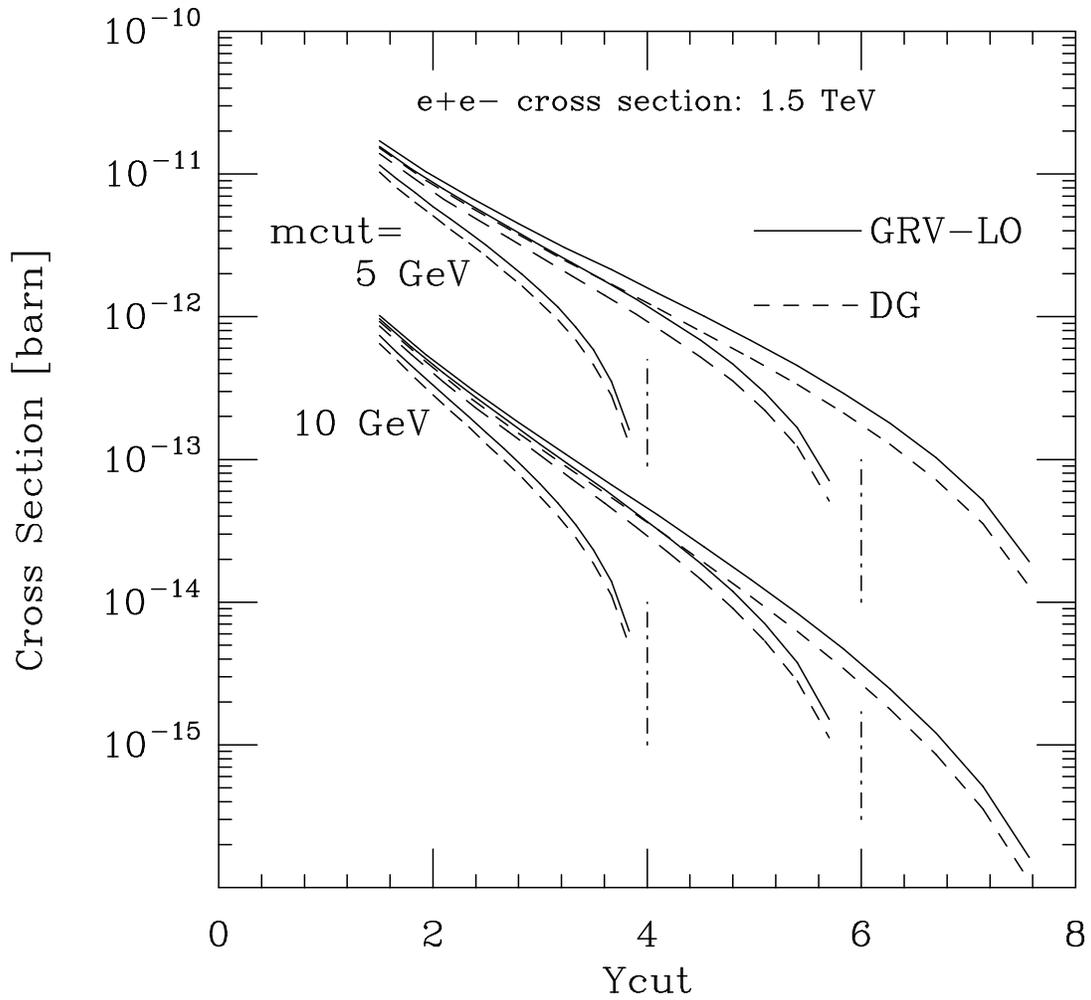}
}
\end{center}
\caption[*]{
            Same as Fig. 5, but at a center-of-mass energy
            of $E_{\rm cm} = 1.5$ TeV.
           }
\label{Fig6}
\end{figure}

With the projected luminosity for the future linear
colliders $\sim 10^{34}$ [cm$^{-2}$s$^{-1}$] ,
the detectors should be sensitive to the physics at
cross section level of $\sim 1$ [fb]. Hence,
unless the detector has very narrow range of rapidity,
gap events with $Y_{\rm cut} = 4$ and $m_{\rm cut} = 5$ GeV
should be produced at an observable rate, this even
when the survival probability is taken into account. Notice that
in going from $0.5$ to $1.5$ TeV, the gap event cross section
increases by about an order of magnitude.
(Naturally, we also have to keep in mind the uncertainty from the
photon structure functions.)

In Ref. \cite{LEPIIGaps}, various mechanisms for the production
of rapidity gap events at LEP-II have been analyzed. These mechanisms
can be characterized as the annihilation of  $e^+e^-$
into two gauge bosons, which subsequently decay into jet pairs.
As opposed to the resolved photon mechanism studied here, in
the annihilation mechanisms all the beam energy goes into
the production of the hadronic jets. In principle, it is possible
to distinguish these two mechanisms, by measuring the
presence or absence of the $e^+e^-$ in the forward or backward
direction. (Calorimetry may also help, although the lepton colliders
are not expected to be sensitive to forward and backward jets due
to the background problems). In practice, this distinction may not
always be feasible. In terms of orders of magnitude, the
annihilation mechanisms like $e^+e^- \to \gamma^*\gamma^* \to jets$ and
$e^+e^- \to \gamma^* Z \to jets$ may be produced at a competing level
with the resolved photon cases (at least for the $0.5$ TeV machine). Also,
there are other mechanisms of producing gap events, such as coming from
$W$ and $Z$ bosons, via annihilation or resolved mechanisms.
In summary, there is a rich phenomenology still to be studied.
We limit our scope here only to the resolved photon contribution,
and postpone a more comprehensive
analysis of rapidity gap jet events at NLC for the future.

\section{Resolved Photon Gap Events at $\gamma\gamma$ Collider}

As opposed to the $e^+e^-$ case, there is no photon flux suppression
for real photon collisions. Therefore, we expect the resolved
photon events to provide a much larger cross section for gap
events.

\begin{figure}[htbp]
\begin{center}
\leavevmode
{
 \epsfxsize=3.00in
 \epsfbox{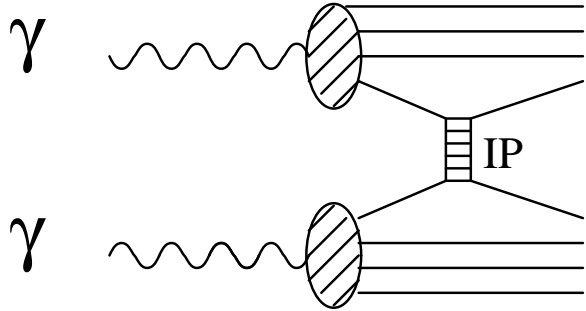}
}
\end{center}
\caption[*]{
            Resolved photon mechanism for producing jet
            events with a large rapidity gap in $\gamma\gamma$
            collision. The partons inside the photons
            undergo a hard scattering via the exchange
            of a perturbative QCD pomeron.
           }
\label{Fig7}
\end{figure}

The production mechanism is illustrated in Fig. 7.
As in the case of $e^+e^-$ collider, we will assume
a rapidity range of $[-\eta_{\rm det}, \eta_{\rm det}]$
for the detector. Although theoretically a $\gamma\gamma$
collider should have little beamstrahlung effects, hence the
detectors should be able to observe jets near the forward
and backward direction, in practice this may not be
true. There remains serious technological challenge
to the conversion of $e^+e^-$ colliders into $\gamma\gamma$
colliders. In particular, the distance between the
laser-backscattering points and the $\gamma\gamma$ collision
point may not be large enough for the deflection of
the remnant $e^+e^-$ beams \cite{Heusch}.

We also consider $0.5$ to $1.5$ TeV
as the range for the center-of-mass energy. The event
cross section is
\begin{equation}
\sigma_{\gamma\gamma} (s,m_{\rm cut}^2,Y_{\rm cut},\eta_{\rm det})
=
\int_{(m_{\rm cut}^2, Y_{\rm cut}, \eta_{\rm det})}
dx_1 \ dx_2 \ dt \
P_{\gamma}(x_1,-t) \ P_{\gamma}(x_2,-t) \
\frac{d\sigma_{gg}}{dt}
\bigl( \hat{s} = x_1 x_2 s, t
\bigr).
\label{EqGammaGammaGapCrossSection}
\end{equation}
And the integration limits for the $t$ variable are similar
to those ones given in Eq.
(\ref{EqLimitA}), (\ref{EqLimitB}) and (\ref{EqLimitC}),
upon substituting $z_1=z_2=1$.

\begin{figure}[htbp]
\begin{center}
\leavevmode
{
 \epsfysize=5.00in
 \epsfbox{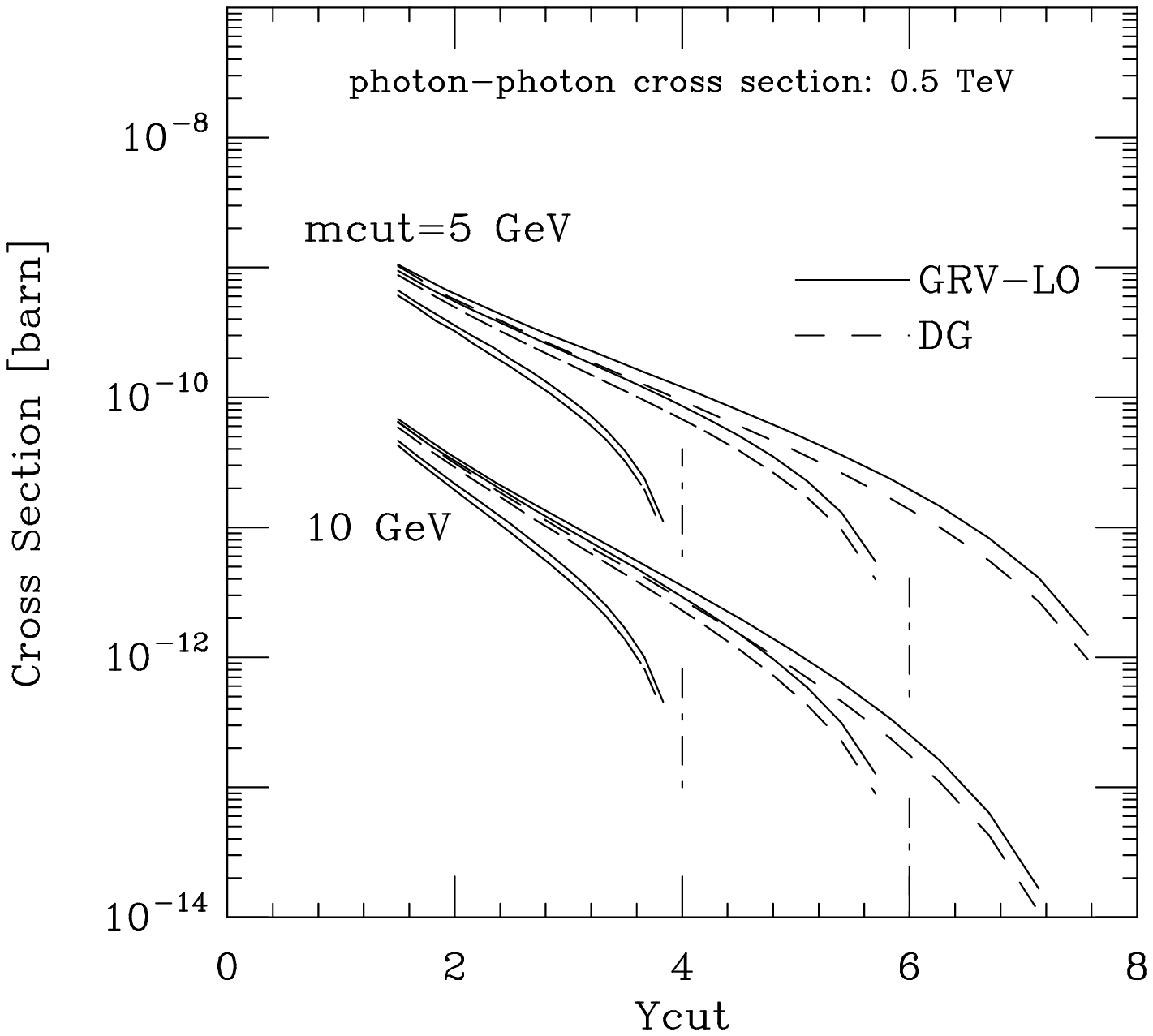}
}
\end{center}
\caption[*]{
            Gap event cross section for $\gamma\gamma$ collider
            at $E_{\rm cm}=0.5$ TeV, for various values
            of transverse momentum cut $m_{\rm cut}=5,10$ GeV
            and detector rapidity limits $\eta_{\rm det}=2,3,4$.
            The maximum observable rapidity gaps are
            $2 \eta_{\rm det} = 4,6,8$ and are indicated by the
            dot-dashed vertical lines. The solid lines are
            obtained by using the Gl\"uck-Reya-Vogt parametrization
            of the photon structure functions, and the broken lines
            by using the Dree-Godbole parametrization.
           }
\label{Fig8}
\end{figure}

\begin{figure}[htbp]
\begin{center}
\leavevmode
{
 \epsfysize=5.00in
 \epsfbox{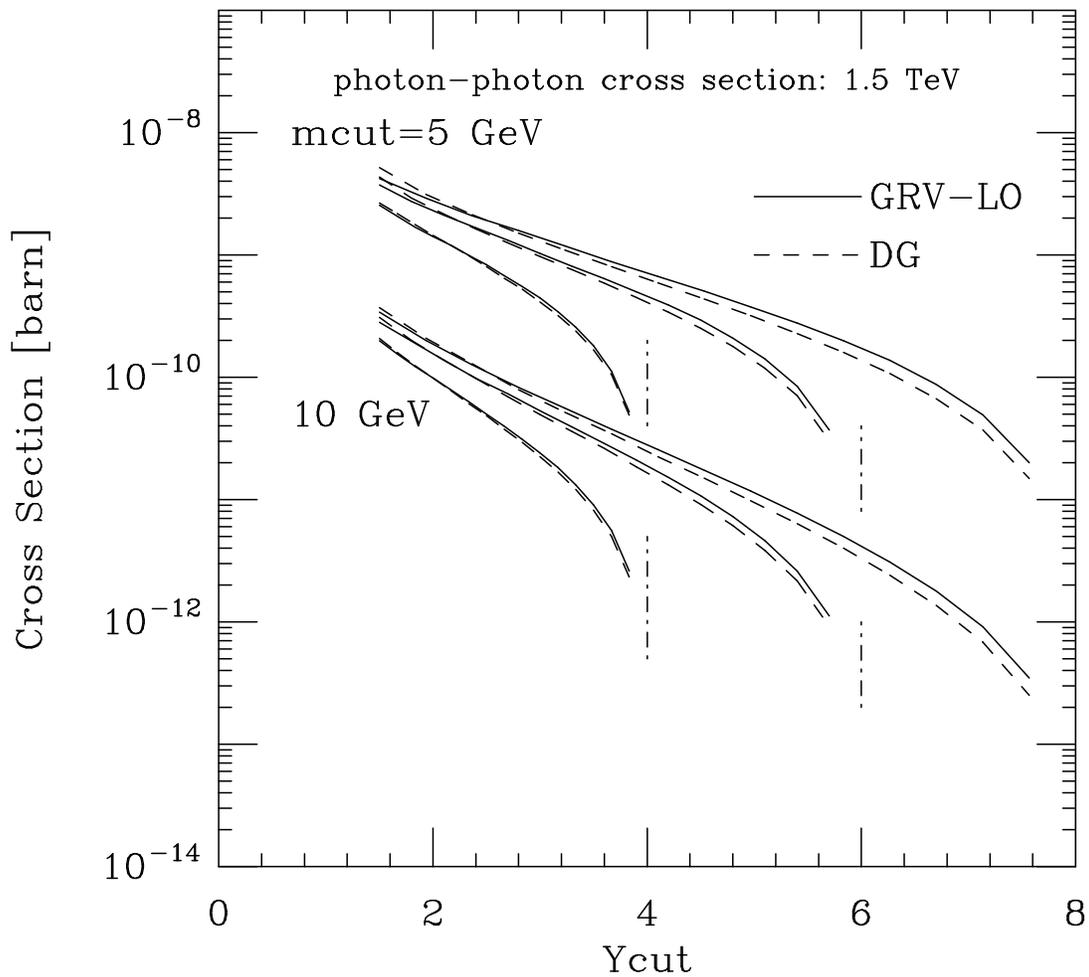}
}
\end{center}
\caption[*]{
            Same as Fig. 8, but at a center-of-mass energy
            of $E_{\rm cm} = 1.5$ TeV.
           }
\label{Fig9}
\end{figure}

The results for the cross sections are plotted in Fig. 8 and
Fig. 9. We can see that compared to the $e^+e^-$ case, the
cross sections are now two to three orders of magnitude larger.
Hence the gap events will be produced copiously at $\gamma\gamma$
colliders. This would provide the ideal environment of the
study of survival probability involving photon initial states.

\section{Conclusions}

We have seen that resolved photons provides a mechanism
of producing jet events containing large rapidity gaps,
and we have analyzed the event cross section at HERA
$ep$ collider and future $e^+e^-$ and $\gamma\gamma$
colliders. We have seen that in all three cases the
event cross section are at the reach of the experiments.
In the case of HERA, it would be interesting to
observe the existence of rapidity gaps between two
hard jet systems (one in the forward direction and
the other one in the backward direction), and analyze
the dependence of the cross section on the rapidity
gap cut $Y_{\rm cut}$ and on the transverse momentum
cut $m_{\rm cut}$. This would provide a first
look into the survival probability involving
photon-hadron collision. In the case of $\gamma\gamma$
collision we have seen that the event cross section
becomes two to three orders of magnitude larger
than the $e^+e^-$ case. We have also seen
that the resolved photon gap events increase
significantly with the total center-of-mass energy.
The observation of these events will allow the study
of the perturbative QCD pomeron physics in environments
alternative to the hadron-hadron
colliders, provide insight into the survival
probability physics of photons, and also allow further
understanding on the relative importance of gap events
coming from random fluctuation of hadronization effects
and from perturbative mechanisms.

We very especially thank Wai-Keung Tang, for all the help
received during the preparation of this work. We also thank
Ina Sarcevic, Stanley Brodsky and Clemens A. Heusch for helpful
conversations, and M. Drees, M. Gl\"uck, E. Reya and A. Vogt
for providing the subroutines for the photon structure functions.

This work was supported by U.S. Department of Energy
Grants No. DE--FG03--93ER40792.

\end{document}